\documentclass[conference]{IEEEtran}
\IEEEoverridecommandlockouts
\usepackage{cite}
\usepackage{amsmath,amssymb,amsfonts}
\usepackage{algorithmic}
\usepackage{tikz}
\usepackage{circuitikz}
\usetikzlibrary{shapes,arrows}
\usepackage{graphicx}
\usepackage{textcomp}
\usepackage{xcolor}
\usepackage[dvipsnames]{xcolor}
\def\BibTeX{{\rm B\kern-.05em{\sc i\kern-.025em b}\kern-.08em
    T\kern-.1667em\lower.7ex\hbox{E}\kern-.125emX}}

\usetikzlibrary{arrows.meta}
\usepackage{acronym}
\usepackage{comment}

\newcommand{\ped}[1]{\ensuremath{_{\mathrm{#1}}}}
\newcommand{\ap}[1]{\ensuremath{^{\mathrm{#1}}}}
\newcommand{\qd}{\ap{qd}}

\acrodef{gfl}[GFL]{grid-following}
\acrodef{gfm}[GFM]{grid-forming}
\acrodef{vsc}[VSC]{voltage source converter}
\acrodef{sg}[SG]{synchronous generator}
\acrodef{pcc}[PCC]{point of common coupling}
\acrodef{ibr}[IBR]{inverter-based resource}
\acrodef{pf}[PF]{participation factor}
\acrodef{scr}[SCR]{short-circuit ratio}
\acrodef{pll}[PLL]{phase-locked loop}
\acrodef{ssa}[SSA]{small-signal stability analyses}
\acrodef{rf}[RF]{rotating reference frame}

\IEEEoverridecommandlockouts

\newcommand\copyrighttext{%
  \footnotesize
  \centering\copyright~2026 IEEE. Personal use of this material is permitted. Permission from IEEE must be obtained for all other uses, in any current or future media, including reprinting/republishing this material for advertising or promotional purposes, creating new collective works, for resale or redistribution to servers or lists, or reuse of any copyrighted component of this work in other works.}
\newcommand\copyrightnotice{%
\begin{tikzpicture}[remember picture,overlay]
\node[anchor=north,yshift=0pt] at (current page.north)
{\setlength{\fboxrule}{0pt}\fbox{\parbox{\dimexpr\textwidth-\fboxsep-\fboxrule\relax}{\copyrighttext}}};
\end{tikzpicture}%
}

\begin{document}

\title{Stability Analysis of Grid-Following and Grid-Forming Converters Connected to Generators\thanks{This work has been funded by the EU fund Next Generation EU, Missione 4, Componente 1, CUP D53D23001650006, MUR PRIN project 2022ZJPPSN SCooPS.}
}
\author
{

\IEEEauthorblockN{Alessandra Casiraghi,
Marzio Barresi, Samuele Grillo}
\\
\IEEEauthorblockA{Dipartimento di Elettronica, Informazione e Bioingegneria, Politecnico di Milano, Milan, Italy.\\
\{alessandra.casiraghi, marzio.barresi, samuele.grillo\}@polimi.it}
}

\IEEEaftertitletext{\copyrightnotice\vspace{0.2\baselineskip}}
\maketitle

\begin{abstract}
This work presents an examination of the main interactions between \ac{gfl} and \ac{gfm} \acp{vsc} and \acp{sg}, capturing the dynamics of a real power grid and pointing out the limitations of considering an ideal one for stability studies. Eigenvalue trajectories and participation factors are studied to perform in-depth small-signal analyses. Specifically, the \ac{gfl} and \ac{gfm} converters are compared in different grid strength scenarios by varying their rating powers and the grid short circuit ratio. Then, time-domain simulations of the non-linear and the developed linear systems are run to validate the mathematical findings from the stability analysis. The results reveal that the stability of \acp{vsc}-dominated grids, either in \ac{gfl} or \ac{gfm} mode, is strongly affected by both the grid strength and the \ac{vsc} power, due to the coupling between the \ac{vsc} control and the \acp{sg}. 
\end{abstract}

\acresetall

\section{Introduction}
The capacity of renewable energy sources has increased rapidly in the last decade and is expected to continue to rise in the near future \cite{intro0}. In this context, the penetration of \acp{ibr} is projected to grow significantly, and the pivotal role of power converters, such as \acp{vsc}, is undeniable \cite{intro1}. 
Currently, converters operate mainly in \ac{gfl} mode and synchronize with the grid by measuring the voltage angle at the \ac{pcc}. Alternatively, \ac{gfm} converters can actively participate in frequency and voltage regulation, guaranteeing the highest \acp{ibr} penetration \cite{gfor1}.

Numerous studies in the literature address the issue of stability in \ac{ibr}-dominated power grids from various perspectives: 
\cite{intro3,intro4,intro5} highlight that grids with a high penetration of \acp{ibr} may experience significant converter-driven stability issues, emphasizing the urgent need for in-depth studies related to increasing \acp{ibr} integration. In particular, some of these challenges have been explored by examining the converter control modes under varying grid strength scenarios, showing that 
the \ac{gfl} converter encounters instability in a weak power grid, while the \ac{gfm} loses stability in a strong power grid \cite{intro6,intro7}; furthermore, the stability boundaries of both control strategies have proved to be highly dependent on proper tuning of regulators \cite{intro8}. However, the majority of existing stability studies, considering different grid strength conditions, have been performed assuming an ideal grid, thereby neglecting the frequency and voltage dynamics of a real power grid. In fact, to the best of our knowledge, few studies have explored the behavior of \acp{vsc} when connected to a \ac{sg}, which comprehensively captures the dynamics of a real power grid relying on conventional generating units, such as thermal or hydro-based generation \cite{linear1}. 

In order to enhance the aforementioned analyses on \acp{vsc} operation in weak and strong grid scenarios, this work includes a detailed investigation of the interactions between \ac{gfl} and \ac{gfm} converters and \acp{sg}, through small-signal analyses based on eigenvalues and \acp{pf}, highlighting the main differences in the interactions and stability margins when considering a real \ac{sg}-based grid instead of an ideal one. Specifically, \ac{gfl} and \ac{gfm} performances are compared under different grid strength conditions by varying the \ac{vsc} power and the grid \ac{scr}, which denotes a strong grid if greater than 3 and a very weak grid if less than 2.


\section{System Modeling}
\label{sec2}
\begin{figure}[b!]
    \centering
    \resizebox{0.43\textwidth}{!}{
    \begin{circuitikz}[font={\Large}]
    \ctikzset{bipoles/diode/height=0.25,bipoles/diode/width=0.25,
              resistors/scale = 1.2,
              resistors/width = 0.2,
              resistors/zigs = 2,
              inductors/scale = 1.2,
              inductors/width=0.3,
              inductors/coils = 3,
              sources/scale = 0.7,
              capacitors/scale = 0.7
            }
    
    \draw[black]
    (4-0.2+0.5,-0.5)   to [short] ++  (0,-0.5) node [nigbt,anchor=D, color=black]  (igbt1) {}   
    (igbt1.S) to [short] ++ (0,-0.5) coordinate (T1-);
    \draw[black] 
    (igbt1.S) -- ++ (0.3,0)  coordinate (D1) to [Do,color=black] (D1 |- igbt1.D) -- (igbt1.D);       
    \draw [rounded corners] (3-0.2+0.5,-2.7-0.5) rectangle (5-0.2+0.5,0.2-0.5);

    \draw (17.25,-1) circle (0.5);
    \node[scale=1.8] at (17.25,-1) {$\sim$};
    \node at (17.25, -.2) {\textcolor{gray}{\textbf{SG}}};
    \draw[->, black, line width = 1.1] (16.4, -.7) to (15.9, -.7);
    \node at (16.2, -.4) {$\textcolor{black}{i\ped{g}}$};
    \draw (16.75,-1)to[short] (15.5-0.5,-1);
    \draw[-] (15.5, -1) to (15.5, -2);
    \draw (15.5,-1) node[circle, fill=black, inner sep = 1pt]{};
    \node at (15.5, -.6) {$\textcolor{black}{v\ped{g}}$};
    
    \draw[-{Triangle[open,line width=1.5pt,width=7mm,length=6mm]}] (15.5, -2) -- (15.5,-3.7);
    \draw[rounded corners, dashed, gray, line width = 0.8mm] (15, -4.1) rectangle (17.2, -2.1);
    \node at (16.4, -3.7) {\textbf{\textcolor{gray}{Load}}};
    \draw[->, black, line width = 1.1] (15.7, -1.3) to (15.7, -1.8);
    \node at (16, -1.6) {$\textcolor{black}{i\ped{l}}$};
    
    \draw (7.5+0.5,-1)to[short] (10-0.2,-1);
    \draw (10-0.2-0.25+3.1,-1) to[R,l^=$R\ped{l}$](10.75 -0.2-0.25+3.1,-1);
    \draw (10.75-0.2-0.25+3.1,-1) to [L,l^=$L\ped{l}$](12.25-0.2+3.1,-1);
    \draw [rounded corners, gray, dashed, line width = 0.8mm] (9.6-0.2+3.1,-2.4) rectangle (11.7+3.1,-.1);
    \node at (10.55+3.1, -2) {\textcolor{gray}{\textbf{Line}}};
    \draw[->, black, line width = 1.1] (10.3+3.1, -1.2) to (10.8+3.1, -1.2);
    \node at (10.55+3.1, -1.5) {$\textcolor{black}{i\ped{t}}$};
    
    \draw (7.5+0.5,-1) node[circle, fill=black, inner sep = 1pt]{};
    \node at (7.5+0.5, -.6) {$\textcolor{black}{v_\mathrm{PCC}}$};
    \draw (7.5+0.5,-1)to[short] (7.5+0.5,-1.5);
    \draw (7.5+0.5,-1.5) to[R,l^=$R\ped{fc}$](7.5+0.5,-2.4);
    \draw (7.5+0.5,-2.4) to [C,l^=$C\ped{f}$](7.5+0.5,-3.4);
    \draw (7.5+0.5,-3.4) node[ground]{};

    \draw[-] (15.5-2.4-0.4, -1) to (14.2-2.4, -1);
    \draw (13-2.4, -1) circle (0.5);
    \draw (13.7-2.4, -1) circle (0.5);
    \draw[-] (12.25-0.2-2.4,-1) to (12.5-2.4, -1);
    \draw [rounded corners, gray, dashed, line width = 0.8mm] (11.9-2.4,-2.4) rectangle (14.7-2.4,-.1);
    \node at (13.3-2.4, -1.9) {\textcolor{gray}{\textbf{Transformer}}};
    
    \draw(4.8+0.5,-1) to[R,l^=$R\ped{f}$](5.8+0.5,-1);
    \draw (5.8+0.5,-1) to [L,l^=$L\ped{f}$](6.8+0.5,-1);
    \draw[->, black, line width = 1.1] (5.6+0.5, -1.2) to (6.1+0.5, -1.2);
    \node at (5.85+0.5, -1.5) {$\textcolor{black}{i\ped{c}}$};
    \draw (6.8+0.5,-1)to[short] (7.5+0.5,-1);
    \draw [rounded corners, gray, dashed, line width = 0.8mm](2.5+0.5, -4.5) rectangle (8.75+0.5, -.1);
    \node at (4+0.5, -4) {\textbf{\textcolor{gray}{VSC}}};
    \end{circuitikz}
}
\caption{Single line diagram of the studied system.}
\label{fig:completo}
\end{figure}
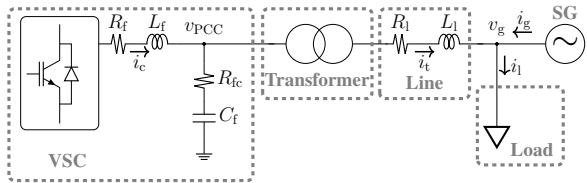

The system in \figurename~\ref{fig:completo} is analyzed in order to study an \acp{ibr}-dominated power grid: it includes an \ac{sg}, which represents thermal-based units, an aggregate model of \acp{vsc}, which can operate both in \ac{gfl} and in \ac{gfm} modes, and a power demand. 
$R\ped{f}$, $R\ped{fc}$ $L\ped{f}$, $C\ped{f}$ are the \ac{vsc} filter resistances, inductance, and capacitance. 
A connection line links the aggregate \ac{vsc} to the \ac{sg}, and it is directly related to the grid strength when the other impedances are fixed: the line length and, in turn, its impedance ($R\ped{l}$, $L\ped{l}$) are determined from a specific value of \ac{scr} through the Thévenin equivalent at the \ac{pcc}. Lastly, the system operates at two voltage levels: the \ac{vsc} and its filter are referred to the transformer LV side (660 V), while the connection line, the load, and the \ac{sg} to the HV side (24 kV).
\subsection{Synchronous Generator}
\begin{figure}[t!]
\centering
\resizebox{0.36\textwidth}{!}{
\ctikzset{
resistors/scale = 1.2,
resistors/width = 0.2,
resistors/zigs = 2,
inductors/scale = 1.2,
inductors/width=0.3,
inductors/coils = 3,
sources/scale = 0.7
}
\begin{circuitikz}[american voltages,
font={\Large}
]

\draw[->, black, line width = 0.2mm] (14.1-2-5+1, 4-1.5+0.2) to (14.6-2-6.4+1, 4-1.5+0.2);
\draw[->, black, line width = 0.2mm] (14.6-2-6.4+1, 4-1.5-0.2) to (14.1-2-5+1, 4-1.5-0.2);
\node at (14.9-2-6.2+1, 4-1.5+0.2+0.2) { {$\textcolor{black}{v\ped{f}}$}};
\node at (14.9-2-6.2+1-0.05, 4-1.5+0.2-0.8) {$v\ped{g}\qd$};
\draw[draw = black] (12.9-2-3.8+1, 3.6-1.5) rectangle(14.1-2-3.8+1+0.4, 4.4-1.5);
\node at (13.5-2-3.8+1+0.2, 4-1.5) [scale = 0.9] {Exciter};
\node at (11.5-2.15+0.4+1-0.4-0.2 + 0.5, 2.5) {\colorbox{gray!40!white}{$V_\mathrm{SG}\ap{ref}$}};
\draw [->, black, line width = 0.2mm] (11.5-2.15+0.4+1-0.4-0.8+0.5, 2.5) to (11.5-2.15+0.4+1-0.4-1.05+0.4, 2.5);

\node at (4.5-1.4-0.4-1, 4-1.5+0.4) { {\colorbox{gray!40!white}{$\omega\ped{n}$}}};
\node at (4.5-1.4-0.4-1, 2.2-0.1) {\colorbox{gray!40!white}{$P\ped{m}\ap{ref}$}};
\draw[->, black, line width = 0.2mm] (4.5-1.4-1,4-1.5+0.4) to (4.9-1.4-1, 4-1.5+0.4); 
\draw[->, black, line width = 0.2mm] (4.5-1.4-0.8,2.1) to (4.9-1.4-1, 2.1); 
\draw[draw = black] (4.9-1.4-1, 3.6-1.5-0.1-0.1) rectangle (6.2-1.4,4.4-1.5+0.1+0.1);
\node at (5.55-1.4-0.5, 4-1.5) [scale = 0.9, align=center] {Governor\\Turbine};
\draw[->, black, line width = 0.2mm] (6.2-1.4,4-1.5+0.3) to (6.8-1.4+0.4, 4-1.5+0.3); 
\draw[->, black, line width = 0.2mm] (6.8-1.4+0.4, 4-1.5-0.3)  to (6.2-1.4,4-1.5-0.3); 
\node at (6.6-1.4, 4.3-1.5+0.3) { {$\textcolor{black}{P\ped{m}}$}};
\node at (6.6-1.4+0.1, 4-1.5-0.55) {$\omega\ped{r}$};

\draw[rounded corners, gray, line width = 0.5 mm] (5.8, 2) rectangle (7.2, 3);
\node at (5.8 + 0.7, 2.5) {SG};
\draw[->, line width = 0.2mm] (6.1,3.4) to (6.1,3);
\node at (6.1, 3.75) {$v\ped{g}\ap{qd}$};
\draw[->, line width = 0.2mm] (6.9,3) to (6.9,3.4);
\node at (6.9, 3.75) {$i\ped{g}\ap{qd}$};

\end{circuitikz}
}
\caption{\ac{sg} equivalent representation.}
\label{fig:SG}
\end{figure}
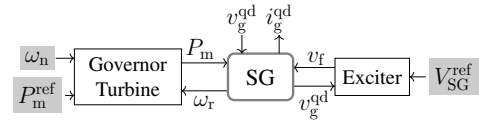

The block diagram of the \ac{sg} model in the rotor reference frame is shown in \figurename~\ref{fig:SG} with its control structures: the frequency regulation system and the excitation system. 

A round-rotor synchronous machine is considered. Its parameters are taken from \cite{sg1}. It includes the $d$ and $q$ stator windings with voltages $v\ped{g}\qd$ and currents $i\ped{g}\qd$, the excitation circuit on the $d$-axis, and three damper windings. Mechanical dynamics is represented as a single mass model with constant moment of inertia, according to the swing equation.

The excitation system, based on the IEEE AC4C standard model in \eqref{eq:exciter}, regulates the voltage at the \ac{sg} terminals by providing the field voltage $v\ped{f}$. $\tau\ped{B}$ and $\tau\ped{C}$ are the time constants of the lead-lag compensator, while $k\ped{A}$ and $\tau\ped{A}$ are the regulator gain and time constant, tuned according to \cite{sg3}. 
\begin{equation}
\label{eq:exciter}
    \mathrm{G}_\mathrm{exc}(s) = \frac{1 + s\tau\ped{C}}{1 + s\tau\ped{B}}\cdot \frac{k\ped{A}}{1 + s\tau\ped{A}}
\end{equation}

The governor and the turbine constitute the frequency regulation system, according to the standard IEEEG1
. The turbine controls the mechanical power $P\ped{m}$ to its reference value $P\ped{m}\ap{ref}$; the simplified transfer function $\mathrm{G}_\mathrm{turb}$ of a single reheat tandem compound turbine is considered in \eqref{eq:turbine}, where $K_\mathrm{HP}$ and $K_\mathrm{LP}$ are the torque fractions of high- and low-pressure turbines, and $T_\mathrm{RH}$ is the reheater time constant \cite{sg1}. The governor, with transfer function $\mathrm{G}_\mathrm{gov}$, controls the shaft rotational speed $\omega\ped{r}$; $k$ is the droop, and $\tau\ped{R}$ is the control valve time constant:
\begin{equation}
\label{eq:turbine}
    \mathrm{G}_\mathrm{turb}(s) = K_\mathrm{HP} + \frac{K_\mathrm{LP}}{1 + sT_\mathrm{RH}}, \quad \mathrm{G}_\mathrm{gov}(s) = \dfrac{k}{1 + s\tau\ped{R}}
\end{equation}
\subsection{Voltage Source Converter}
As this paper aims to perform small-signal analyses, an average model is used, neglecting the switching. The AC port is modeled as a voltage source, controlled by the \ac{vsc} control scheme. The DC port is decoupled from the AC side, and its dynamics is influenced by power exchanges with the AC grid. If not controlled, it is modeled as a constant voltage source, otherwise as a constant current source with a DC-link capacitor $C\ped{c}$. To study high \acp{ibr} penetration scenarios, an aggregate model of $N$ \acp{vsc} is used, following the procedures outlined in \cite{VSC1} and \cite{VSC2} for \ac{gfl} and \ac{gfm}, respectively. 
\subsubsection{Grid-following converter}
GFL mode consists in injecting active and reactive power into an energized grid, with the \ac{vsc} acting as a current source
\cite{gfol2}. 
The \ac{pll} plays a key role in guaranteeing its proper operation: it ensures grid synchronization by tracking frequency and angle at \ac{pcc} and providing the reference frame for the control loops. 

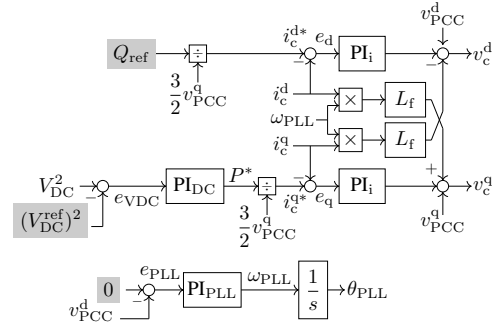
\begin{figure}[t!]
    \centering
    \resizebox{0.37\textwidth}{!}{
    \begin{circuitikz}
        \draw (0, 0.2+0.6) rectangle (0.8, 0.8+0.6);
        \node at (0.4, 0.5+0.6) {PI$\ped{i}$};
        \draw[->] (-0.4, 0.5+0.6) to (-0.0, 0.5+0.6);
        \draw (-0.5, 0.5+0.6) circle (0.1cm);
        \node at (-0.7, 0.75+0.6-0.1) [scale = 0.8]{$-$};
        \draw[->] (-1.1, 0.5+0.6) to (-0.6, 0.5+0.6);
        \draw[->] (-0.5, 1.8) to (-0.5, 0.6+0.6);
        \draw[->] (-0.8, 1.8) to (0, 1.8);
        \node at (-1, 1.8) {$i\ped{c}\ap{q}$}; 
        \draw (0, 1.7) rectangle (0.4, 2.1);
        \node at (0.2, 1.9) {$\times$};
        \draw[->] (0.4, 1.9) to (0.8, 1.9);
        \draw (0.8, 1.6) rectangle (1.5, 2.2);
        \node at (1.15, 1.9) {$L\ped{f}$};
        \draw[-] (1.5, 1.9) to (1.6, 1.9);
        \draw[-] (1.6, 1.9) to (1.8, 2.4);
        \draw[->] (1.8, 2.4) to (1.8, 3.9-0.6);
        \draw (1.8, 4-0.6) circle(0.1cm);
        \node at (1.6, 3.8-0.6) [scale = 0.8] {$-$};
        \draw[->] (.8, 4-0.6) to (1.7, 4-0.6);
        \draw[->] (1.8, 4.45-0.6) to (1.8, 4.15-0.6);
        \node at (1.8, 4.7-0.6) {$v_\mathrm{PCC}\ap{d}$};
        \draw[->] (1.9, 4-0.6) to (2.3,4-0.6);
        \node at (2.5, 4-0.6) {$\textcolor{black}{v\ped{c}\ap{d}}$};
        \node at (-0.25, 0.25+0.6) {$e\ped{q}$};

        \draw (-1.2-1.8, 0.2+0.6) rectangle (-0.2-1.8, 0.8+0.6);
        \node at (-0.7-1.8, 0.5+0.6) {PI$_\mathrm{DC}$};
        \draw[->] (-2.2-1.8, 0.5+0.6) to (-1.2-1.8, 0.5+0.6);
        \node at (-1.7-1.8, 0.25+0.6) {$e_\mathrm{VDC}$};
        \draw (-2.3-1.8, 0.5+0.6) circle(0.1cm);
        \draw[->] (-2.7-1.8,0.5+0.6) to (-2.4-1.8, 0.5+0.6);
        \node at (-3.1-1.8, 0.5+0.6) {$\textcolor{black}{V_\mathrm{DC}^2}$};
        
        \draw[-] (-2.5-1.8, -0.2+0.6) to (-2.3-1.8, -0.2+0.6);
        \draw[->] (-2.3-1.8, -0.2+0.6) to (-2.3-1.8, 0.4+0.6);
        \node at (-3-1.8-0.2, -2 + 1.9+0.6) {\colorbox{gray!40!white}{$(V_\mathrm{DC}\ap{ref})^2$}};
        \node at (-2.5-1.8, -1.6+1.9+0.6) [scale = 0.8] {$-$};

        \draw[->] (-0.2-1.8, 0.5-1.7+1.7+0.6) to (0.4-1.8, 0.5-1.7+1.7+0.6);
        \node at (0.1-1.8, 0.7-1.7+1.7+0.6) {$P^*$};
        \draw (0.4-1.8, 0.35-1.7+1.7+0.6) rectangle (0.7-1.8, 0.65-1.7+1.7+0.6);
        \node at (0.55-1.8, 0.5-1.7+1.7+0.6) [scale = 0.7] {$\div$};
        \draw[->] (0.55-1.8, 0-1.7+1.7+0.6) to (0.55-1.8, 0.35-1.7+1.7+0.6);
        \node at (0.55-1.8, -0.2-1.7+1.7+0.6) {$\dfrac{3}{2}v_\mathrm{PCC}\ap{q}$};
        \node at (1.25-2, 0.5-1.7+1.4+0.6) {$i\ped{c}\ap{q*}$};

        \node at (-0.8, 2.25) {\textcolor{ black!70!black}{$\omega_\mathrm{PLL}$}};
        \draw[-] (-0.4, 2.25) to (-0.2, 2.25);
        \draw[-] (-0.2, 2.25) to (-0.2, 2);
        \draw[->] (-0.2, 2) to (0, 2);
        \draw[-] (-0.2, 2.25) to (-0.2, 2.5);
        \draw[->] (-0.2, 2.5) to (0, 2.5);
        \draw [draw =  black!70!black] (-5.5+2+0.8, 2+2.5-6+0.5) rectangle (-4.5+2+0.8, 2.6+2.5-6+0.5);
        \node at (-5+2+0.8, 2.3+2.5-6+0.5) {\textcolor{ black!70!black}{PI$_\mathrm{PLL}$}};
        \draw[->,  black!70!black] (-4.5+2+0.8, 2.3+2.5-6+0.5) to (-3.5+2+0.8, 2.3+2.5-6+0.5);
        \node at (-4+2+0.8, 2.5+2.5-6+0.5) {$\textcolor{ black!70!black}{\omega_\mathrm{PLL}}$};
        \draw[draw =  black!70!black] (-3.5+2+0.8, 1.8+2.5-6+0.5) rectangle (-3+2+0.8, 2.8+2.5-6+0.5);
        \node at (-3.25+2+0.8, 2.3+2.5-6+0.5) {$\textcolor{ black!70!black}{\dfrac{1}{s}}$};
        \draw[->,  black!70!black] (-3+2+0.8, 2.3+2.5-6+0.5) to (-2.7+2 +0.8, 2.3+2.5-6+0.5);
        \node at (-2.3+2+0.8, 2.3+2.5-6+0.5) {$\textcolor{ black!70!black}{\theta_\mathrm{PLL}}$};
        \draw[->,  black!70!black] (-5.9+2-0.1+0.8, 2.3+2.5-6+0.5) to (-5.5+2+0.8, 2.3+2.5-6+0.5);
        \node at (-5.9+2 + 0.8, 2.6+2.5-6+0.5) {$\textcolor{ black!70!black}{e_\mathrm{PLL}}$};
        \draw (-6.1+2 + 0.8, 2.3+2.5-6+0.5) circle(0.1cm);
        \draw[->] (-6.5+2 + 0.8, 2.3+2.5-6+0.5) to (-6.2+2 + 0.8, 2.3+2.5-6+0.5);
        \node at (-6.6+2-0.2 + 0.8, 2.3+2.5-6+0.5) {\colorbox{gray!40!white}{$0$}};
        \draw[->] (-6.1+2 + 0.8, 1.8+2.5-6+0.5) to (-6.1+2 + 0.8, 2.2+2.5-6+0.5);
        \node at (-6.1+2 + 0.8 - 1, 1.6+2.5-6+0.5+0.35) {$v_\mathrm{PCC}\ap{d}$};
        \node at (-6.3+2 + 0.8, 2.1+2.5-6+0.5) [scale = 0.7] {$-$};
        \draw[-] (-6.1+2 + 0.8 - 0.5, 1.8+2.5-6+0.5) to (-6.1+2 + 0.8, 1.8+2.5-6+0.5);

        \draw (0, 3.7-0.6) rectangle (0.8, 4.3-0.6);
        \node at (0.4, 4-0.6) {PI$\ped{i}$};
        \draw[->] (-0.4, 4-0.6) to (-0.0, 4-0.6);
        \draw (-0.5, 4-0.6) circle (0.1cm);
        \node at (-0.7, 3.75-0.6+0.1) [scale = 0.8] {$-$};
        \draw[->] (-1-1.3, 4-0.6) to (-0.6, 4-0.6);
        \draw[->] (-0.5, 2.7) to (-0.5, 3.9-0.6);
        \draw[->] (-0.8, 2.7) to (0, 2.7);
        \node at (-1, 2.7) {$i\ped{c}\ap{d}$};
        \draw (0, 2.4) rectangle (0.4, 2.8);
        \node at (0.2, 2.6) {$\times$};
        \draw[->] (0.4, 2.6) to (0.8, 2.6);
        \draw (0.8, 2.3) rectangle (1.5, 2.9);
        \node at (1.15, 2.6) {$L\ped{f}$};
        \draw[-] (1.5, 2.6) to (1.6, 2.6);
        \draw[-] (1.6, 2.6) to (1.8, 2.1);
        \draw[->] (1.8, 2.1) to (1.8, 0.6+0.6);
        \draw (1.8, 0.5+0.6) circle (0.1 cm);
        \node at (1.6, 0.8+0.6) [scale=0.8] {$+$};
        \draw[->] (0.8, 0.5+0.6) to (1.7, 0.5+0.6);
        \draw[->] (1.8, -0+0.6) to (1.8, 0.4+0.6); 
        \node at (1.8, -0.1+0.6) {$v_\mathrm{PCC}\ap{q}$};
        \draw[->] (1.9, 0.5+0.6) to (2.3,0.5+0.6);
        \node at (2.5, 0.5+0.6) {$\textcolor{black}{v\ped{c}\ap{q}}$};
        \node at (-0.75, 4.3-0.6) {$i\ped{c}\ap{d*}$};
        \node at (-0.25, 4.25-0.6) {$e\ped{d}$};
        
        \draw[->] (-0.2-1.8-1.2, 4-0.6) to (0.4-1.8-1.2, 4-0.6);
        \node at (0.1-1.8-1.2-0.7, 4-0.6) {\colorbox{gray!40!white}{$Q\ped{ref}$}};
        \draw (0.4-1.8-1.2, 4-0.6-0.15) rectangle (0.7-1.8-1.2, 4-0.6+0.15);
        \node at (0.55-1.8-1.2, 4-0.6) [scale = 0.7] {$\div$};
        \draw[->] (0.55-1.8-1.2, 4-0.6-0.15-0.35) to (0.55-1.8-1.2, 4-0.6-0.15);
        \node at (0.55-1.8-1.2, 4-0.6-0.15-0.35-0.2) {$\dfrac{3}{2}v_\mathrm{PCC}\ap{q}$};

    \end{circuitikz}
    }
    \caption{Grid-following control scheme.}
    \label{fig:GFOL}
\end{figure}

Together with the \ac{pll}, the \ac{gfl} vector control strategy consists of fast current control loops and outer computations to set the reference currents, as shown in \figurename~\ref{fig:GFOL}.
The \ac{pll} is modeled as a PI regulator, tuned by setting the time constant and damping \cite{gfol3}, which determines the frequency $\omega_\mathrm{PLL}$ as output. 
Then, two PI controllers ensure that the \ac{vsc} current $i\ped{c}\ap{qd}$ follows its reference, by providing the converter voltages $v\ped{c}\ap{qd}$ as output; the gains are tuned by fixing the loop time constant. 
Lastly, the DC voltage PI regulator, with time constant $\tau_\mathrm{DC}$, acts on $V_\mathrm{DC}^2$ 
, providing the reference active power and, consequently, the reference $q$-axis current. It is tuned on the basis of the DC voltage open-loop transfer function, computed from the DC power expression. The reference $d$-axis current is computed directly from the reference reactive power $Q\ped{ref}$. 

\subsubsection{Grid-forming converter}
The \ac{vsc} can also operate in \ac{gfm} mode, acting as a voltage source by actively participating in both voltage and frequency regulation \cite{gfor1}. Unlike \ac{gfl} converters that follow the grid frequency through the \ac{pll}, \ac{gfm} converters can directly generate their frequency and phase angle with different strategies, such as virtual synchronous machine, synchronverter, and droop control \cite{gfor5}. 

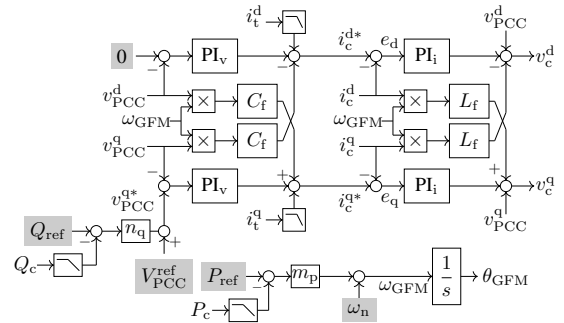
\begin{figure}[t!]
    \centering
    \resizebox{0.415\textwidth}{!}{
    \begin{circuitikz}
        \draw (0, 0.2+0.6) rectangle (0.8, 0.8+0.6);
        \node at (0.4, 0.5+0.6) {PI$\ped{i}$};
        \draw[->] (-0.4, 0.5+0.6) to (-0.0, 0.5+0.6);
        \draw (-0.5, 0.5+0.6) circle (0.1cm);
        \node at (-0.7, 0.7+0.6) [scale = 0.8]{$-$};
        \draw[->] (-1-0.8-0.05, 0.5+0.6) to (-0.6, 0.5+0.6);
        \draw[->] (-0.5, 1.8) to (-0.5, 0.6+0.6);
        \draw[->] (-0.8, 1.8) to (0, 1.8);
        \node at (-1, 1.8) {$i\ped{c}\ap{q}$}; 
        \draw (0, 1.7) rectangle (0.4, 2.1);
        \node at (0.2, 1.9) {$\times$};
        \draw[->] (0.4, 1.9) to (0.8, 1.9);
        \draw (0.8, 1.6) rectangle (1.5, 2.2);
        \node at (1.15, 1.9) {$L\ped{f}$};
        \draw[-] (1.5, 1.9) to (1.6, 1.9);
        \draw[-] (1.6, 1.9) to (1.8, 2.4);
        \draw[->] (1.8, 2.4) to (1.8, 3.9-0.6);
        \draw (1.8, 4-0.6) circle(0.1cm);
        \node at (1.6, 3.8-0.6) [scale = 0.8] {$-$};
        \draw[->] (.8, 4-0.6) to (1.7, 4-0.6);
        \draw[->] (1.8, 4.45-0.6) to (1.8, 4.1-0.6);
        \node at (1.8, 4.7-0.6) {$v_\mathrm{PCC}\ap{d}$};
        \draw[->] (1.9, 4-0.6) to (2.3,4-0.6);
        \node at (2.5, 4-0.6) {$\textcolor{black}{v\ped{c}\ap{d}}$};

        \node at (-0.95, 0.2+0.6) {$i\ped{c}\ap{q*}$};
        \node at (-0.25, 0.25+0.6) {$e\ped{q}$};

        \draw (0-3.75, 0.2+0.6) rectangle (0.8-3.75, 0.8+0.6);
        \node at (0.4-3.75, 0.5+0.6) {PI$\ped{v}$};
        \draw[->] (-0.4-3.75, 0.5+0.6) to (-0-3.75, 0.5+0.6);
        \draw (-0.5-3.75, 0.5+0.6) circle (0.1cm);
        \node at (-0.7-3.75, 0.75+0.6) [scale = 0.8]{$-$};
        \draw[->] (-3.75-0.5, 0.5-0.2+0.1) to (-0.5-3.75, 0.5+0.5);
        \draw[-] (-3.75-0.75, 0.5-0.2) to (-3.75-0.5-0.1, 0.5-0.2);
        \node at (-4.8, 0.25+0.6) {$v_\mathrm{PCC}\ap{q*}$};
        \draw (-3.75-0.5, 0.5-0.2) circle (0.1cm);
        \node at (-0.3-3.75, 0.5-0.4) [scale = 0.8]{$+$};
        \node at (-0.5-3.75, 0.5- 0.8-0.2) {\colorbox{gray!40!white}{$V_\mathrm{PCC}\ap{ref}$}};
        \draw[->] (-3.75-0.5, 0.5-0.6) to (-3.75-0.5, 0.5 - 0.3);
        \draw[->] (-0.5-3.75, 1.8) to (-0.5-3.75, 0.6+0.6);
        \draw[->] (-0.8-3.75, 1.8) to (0-3.75, 1.8);
        \node at (-1-3.95, 1.8) {$v_\mathrm{PCC}\ap{q}$}; 
        \draw (0-3.75, 1.7) rectangle (0.4-3.75, 2.1);
        \node at (0.2-3.75, 1.9) {$\times$};
        \draw[->] (0.4-3.75, 1.9) to (0.8-3.75, 1.9);
        \draw (0.8-3.75, 1.6) rectangle (1.5-3.75, 2.2);
        \node at (1.15-3.75, 1.9) {$C\ped{f}$};
        \draw[-] (1.5-3.75, 1.9) to (1.6-3.75, 1.9);
        \draw[-] (1.6-3.75, 1.9) to (1.8-3.75, 2.4);
        \draw[->] (1.8-3.75, 2.4) to (1.8-3.75, 3.9-0.6);
        \draw (1.8-3.75, 4-0.6) circle(0.1cm);
        \node at (1.6-3.75, 3.8-0.6) [scale = 0.8] {$-$};
        \draw[->] (.8-3.75, 4-0.6) to (1.7-3.75, 4-0.6);
        \draw[->] (1.8-3.75, 4.4-0.6) to (1.8-3.75, 4.1-0.6);
        \draw (1.6-3.75, 4.4-0.6) rectangle(2 - 3.75, 4.8-0.6);
        \draw[-] (1.8 - 3.75-0.15, 4.1) to (1.8-3.75 + 0.05, 4.1);
        \draw[-] (1.8-3.75+0.05, 4.1) to (1.8-3.75+0.15, 3.9);
        \draw[->] (1.8-3.75-0.3-0.2, 4.9-0.6-.3) to (1.8-3.75-0.2, 4.9-0.6-0.3);
        \node at (1.8-3.75-0.7, 4.9-0.6-.2) {$i\ped{t}\ap{d}$};

        \draw (-5.8+0.8, 0.3-0.2) rectangle (-5.3+0.8, 0.7-0.2);
        \node at (-5.55+0.8, 0.5-0.2) {$n\ped{q}$};
        \draw[->] (-6.25+0.9, 0.5-0.2) to (-5.8+0.8, 0.5-0.2); 
        \draw (-6.4+1-0.05, 0.5-0.2) circle(0.1cm);
        \node at (-6.6+1-0.05, 0.3-0.2) [scale = 0.8] {$-$};
        \draw[->] (-6.9+1, 0.5-0.2) to (-6.5+1-0.05, 0.5-0.2);
        \node at (-7.3+1, 0.5-0.2) {\colorbox{gray!40!white}{$Q\ped{ref}$}};
        \draw[->] (-6.4+1-0.05, -0.1-0.2) to (-6.4+1-0.05, 0.4-0.2);
        \draw[-] (-6.7+1, -0.1-0.2) to (-6.4+1-0.05, -0.1-0.2);
        \draw (-7.4+1.2, -0.35-0.2+0.05) rectangle (-6.9+1.2, 0.15-0.2-0.05);
        \draw[-] (-7.15+1.0, -0.2) to (-7.15+1.2, -0.2);
        \draw[-] (-7.15+1.2, -0.2) to (-7.15+1.4, -0.4);
        \draw[->] (-8+1.5, -0.1-0.2) to (-7.4+1.2, -0.1-0.2);
        \node at (-7.7+1, -0.1-0.2) {$Q\ped{c}$};

        \node at (-0.8, 2.25) {\textcolor{black}{$\omega_\mathrm{GFM}$}};
        \draw[-] (-0.4, 2.25) to (-0.2, 2.25);
        \draw[-] (-0.2, 2.25) to (-0.2, 2);
        \draw[->] (-0.2, 2) to (0, 2);
        \draw[-] (-0.2, 2.25) to (-0.2, 2.5);
        \draw[->] (-0.2, 2.5) to (0, 2.5);
        \node at (-0.8-3.75, 2.25) {\textcolor{black}{$\omega_\mathrm{GFM}$}};
        \draw[-] (-0.4-3.75, 2.25) to (-0.2-3.75, 2.25);
        \draw[-] (-0.2-3.75, 2.25) to (-0.2-3.75, 2);
        \draw[->] (-0.2-3.75, 2) to (0-3.75, 2);
        \draw[-] (-0.2-3.75, 2.25) to (-0.2-3.75, 2.5);
        \draw[->] (-0.2-3.75, 2.5) to (0-3.75, 2.5);

        \draw [draw = black] (-5-1+2+2, 2+4-.5+0.1-7+0.7) rectangle (-4.5-1+2+2, 2.6+4-.5-0.1-7+0.7);
        \node at (-4.75-1+2+2, 2.3+4-.5-7+0.7) {\textcolor{black}{$m\ped{p}$}};
        \draw (-2.5-0.1-0.2+2, 2.3+4-.5-7+0.7) circle(0.1cm);
        \draw[->] (-2.8+2, 2.3+4-.5-7-0.4+0.7) to (-2.8+2, 2.3+4-.5-7-0.1+0.7);
        \node at (-2.8+2, 2.3+4-.5-7-0.6+0.7) {\colorbox{gray!40!white}{$\omega\ped{n}$}};
        \draw[->] (-4.5-1+2+2,  2.3+4-.5-7+0.7) to (-2.7-0.2+2, 2.3+4-.5-7+0.7);
        \draw[->, black] (-4-1+2+0.5-0.2+2, 2.3+4-.5-7+0.7) to (-3.5-1+2+1+2, 2.3+4-.5-7+0.7);
        \node at (-4-1+2+1+2, 2+4-.5+0.1-7+0.7) {$\textcolor{black}{\omega_\mathrm{GFM}}$};
        \draw[draw = black] (-3.5-1+2+1+2, 1.8+4-.5-7+0.7) rectangle (-3-1+2+1+2, 2.8+4-.5-7+0.7);
        \node at (-3.25-1+2+1+2, 2.3+4-.5-7+0.7) {$\textcolor{black}{\dfrac{1}{s}}$};
        \draw[->, black] (-3-1+2+1+2, 2.3+4-.5-7+0.7) to (-2.7-1+2+1+2, 2.3+4-.5-7+0.7);
        \node at (-2.3-1+2+0.1+1+2, 2.3+4-.5-7+0.7) {$\textcolor{black}{\theta_\mathrm{GFM}}$};
        \draw[->, black] (-5.4-1+2+0.1+2, 2.3+4-.5-7+0.7) to (-5-1+2+2, 2.3+4-.5-7+0.7);
        \draw (-6.1+0.5-1+2+0.2+2, 2.3+4-.5-7+0.7) circle(0.1cm);
        \draw[->] (-6-1+2+0.2+2, 2.3+4-.5-7+0.7) to (-6.2-1+0.5+2+0.2+2, 2.3+4-.5-7+0.7);
        \node at (-6.6-1+0.5+2-0.1+2, 2.3+4-.5-7+0.7) {\colorbox{gray!40!white}{$P\ped{ref}$}};
        \draw[->] (-6.1-1+0.5+2+0.2+2, 1.7+4-.5-7+0.7) to (-6.1-1+0.5+2+0.2+2, 2.2+4-.5-7+0.7);
        \draw[-] (-10.1+3+2+0.5+2, 1.7+4-.5-7+0.7) to (-6.1-3.5+3+2+0.2+2, 1.7+4-.5-7+0.7);
        \node at (-6.3-1+0.5+2+0.2+2, 2.1+4-.5-7+0.7) [scale = 0.7] {$-$};
        \draw (-10.6+3+2+0.5+2, 1.45+4-.5-7+0.7+0.05) rectangle(-10.1+3+2+0.5+2, 1.95+4-0.5-7+0.7-0.05);
        \draw[-] (-10.35+3+2+0.5+2-0.2, 1.7+4-.5-7+0.7+0.1) to (-10.35+3+2+0.5+2, 1.7+4-.5-7+0.7+0.1); 
        \draw[-] (-10.35+3+2+0.5+2, 1.7+4-.5-7+0.7+0.1) to (-10.35+3+2+0.5+2+0.2, 1.7+4-.5-7+0.7+0.1-0.2);
        \draw[->] (-11+3+2+0.6+2, 1.7+4-.5-7+0.7) to (-10.6+3+2+0.5+2, 1.7+4-.5-7+0.7);
        \node at (-10.8+3+2+0.2+2, 1.5+4-0.5+0.2-7+0.7) {$P\ped{c}$};

        \draw (0, 3.7-0.6) rectangle (0.8, 4.3-0.6);
        \node at (0.4, 4-0.6) {PI$\ped{i}$};
        \draw[->] (-0.4, 4-0.6) to (-0., 4-0.6);
        \draw (-0.5, 4-0.6) circle (0.1cm);
        \node at (-0.7, 3.8-0.6)[scale = 0.8] {$-$};
        \draw[->] (-1-0.8-0.05, 4-0.6) to (-0.6, 4-0.6);
        \draw[->] (-0.5, 2.7) to (-0.5, 3.9-0.6);
        \draw[->] (-0.8, 2.7) to (0, 2.7);
        \node at (-1, 2.7) {$i\ped{c}\ap{d}$};
        \draw (0, 2.4) rectangle (0.4, 2.8);
        \node at (0.2, 2.6) {$\times$};
        \draw[->] (0.4, 2.6) to (0.8, 2.6);
        \draw (0.8, 2.3) rectangle (1.5, 2.9);
        \node at (1.15, 2.6) {$L\ped{f}$};
        \draw[-] (1.5, 2.6) to (1.6, 2.6);
        \draw[-] (1.6, 2.6) to (1.8, 2.1);
        \draw[->] (1.8, 2.1) to (1.8, 0.6+0.6);
        \draw (1.8, 0.5+0.6) circle (0.1 cm);
        \node at (1.6, 0.8+0.6) [scale=0.8] {$+$};
        \draw[->] (0.8, 0.5+0.6) to (1.7, 0.5+0.6);
        \draw[->] (1.8, -0+0.6) to (1.8, 0.4+0.6); 
        \node at (1.8, -0.1+0.6) {$v_\mathrm{PCC}\ap{q}$};
        \draw[->] (1.9, 0.5+0.6) to (2.3,0.5+0.6);
        \node at (2.5, 0.5+0.6) {$\textcolor{black}{v\ped{c}\ap{q}}$};

        \node at (-0.95, 4.3-0.6) {$i\ped{c}\ap{d*}$};
        \node at (-0.25, 4.25-0.6) {$e\ped{d}$};

        \draw (0-3.75, 3.7-0.6) rectangle (0.8-3.75, 4.3-0.6);
        \node at (0.4-3.75, 4-0.6) {PI$\ped{v}$};
        \draw[->] (-0.4-3.75, 4-0.6) to (-0.0-3.75, 4-0.6);
        \draw (-0.5-3.75, 4-0.6) circle (0.1cm);
        \node at (-5, 4-0.6) {\colorbox{gray!40!white}{0}};
        \node at (-0.7-3.75, 3.75-0.6)[scale = 0.8] {$-$};
        \draw[->] (-1-3.75, 4-0.6) to (-0.6-3.75, 4-0.6);
        \draw[->] (-0.5-3.75, 2.7) to (-0.5-3.75, 3.9-0.6);
        \draw[->] (-0.8-3.75, 2.7) to (0-3.75, 2.7);
        \node at (-1-3.95, 2.7) {$v_\mathrm{PCC}\ap{d}$};
        \draw (0-3.75, 2.4) rectangle (0.4-3.75, 2.8);
        \node at (0.2-3.75, 2.6) {$\times$};
        \draw[->] (0.4-3.75, 2.6) to (0.8-3.75, 2.6);
        \draw (0.8-3.75, 2.3) rectangle (1.5-3.75, 2.9);
        \node at (1.15-3.75, 2.6) {$C\ped{f}$};
        \draw[-] (1.5-3.75, 2.6) to (1.6-3.75, 2.6);
        \draw[-] (1.6-3.75, 2.6) to (1.8-3.75, 2.1);
        \draw[->] (1.8-3.75, 2.1) to (1.8-3.75, 0.6+0.6);
        \draw (1.8-3.75, 0.5+0.6) circle (0.1 cm);
        \node at (1.6-3.75, 0.7+0.6) [scale=0.8] {$+$};
        \draw[->] (0.8-3.75, 0.5+0.6) to (1.7-3.75, 0.5+0.6);
        \draw[->] (1.8-3.75, 0.1+0.6) to (1.8-3.75, 0.4+0.6); 
        \draw (1.6-3.75, 0.1+0.6) rectangle(2 - 3.75, -0.3+0.6);
        \draw[-] (1.8 - 3.75-0.15, 4.1-3.5) to (1.8-3.75 + 0.05, 4.1-3.5);
        \draw[-] (1.8-3.75+0.05, 4.1-3.5) to (1.8-3.75+0.15, 3.9-3.5);
        \draw[->] (1.8-3.75-0.5, -0.6+0.6+ 0.5) to (1.8-3.75-0.5+0.3, -0.6+0.6+ 0.5);
        \node at (1.8-3.75-0.7, -0.6+0.6+0.5) {$i\ped{t}\ap{q}$};
            
    \end{circuitikz}
    }
    \caption{Grid-forming control scheme.}
    \label{fig:GFOR}
\end{figure}

The \ac{gfm} control scheme adopted in this paper, in \figurename~\ref{fig:GFOR}, consists of the current control loop, which is the same as in \ac{gfl}, the voltage control loop, and the outer droop control. 
Unlike the \ac{gfl}, here the DC voltage is not controlled. 
Two PI controllers, tuned according to \cite{gfor5}, regulate the voltage at \ac{pcc} to its reference $v_\mathrm{PCC}\ap{qd*}$ by providing the reference currents as output. 
Lastly, the \ac{gfm} frequency $\omega_\mathrm{GFM}$ derives from the active power synchronization loop, while the reactive power droop provides the reference $q$-axis \ac{pcc} voltage; their droop coefficients $m\ped{p}$ and $n\ped{q}$ are set according to the grid code.
\section{Small-Signal Stability Analysis}
\label{sec4}
In this section, the performances of \ac{gfl} and \ac{gfm} \acp{vsc} connected to \ac{sg}-based or ideal power grids are compared using \ac{ssa} based on eigenvalues and \acp{pf}. Different ratings of the aggregate \ac{vsc} are considered, keeping the \ac{sg} power constant: \ac{vsc} power is set to $50\%$, $100\%$ and $150\%$ of \ac{sg} power; the load absorbs $75\%$ of the total installed power, and the \ac{gfl} and \ac{gfm} \acp{vsc} operate at the same steady-state equilibrium point. Then, for each scenario, the grid \ac{scr} is varied from $3.5$ to $1.5$ with steps of $0.1$, resulting in variations in the connection line impedance (more details about the system are available on GitHub\footnote{https://github.com/AlessandraCasiraghi/SSA-GFL-and-GFM-VSCs}). 

To perform \ac{ssa}, the state-space linear model of the complete system is derived, considering different \acp{rf}: the grid and the \ac{sg} align with the rotor, the \ac{gfl} control with the \ac{pcc} voltage, and the \ac{gfm} control with the angle imposed by the synchronization loop; as widely recognized, transitions between \acp{rf} are possible using nonlinear rotation matrices, which depend on their angle difference $\Delta\theta$. 

In the following, eigenvalues are displayed in different colors according to the variables associated with sufficiently high \ac{pf} ($>30\%$) \cite{linear1}. Thus, the color legend indicates the state variables significantly participating in each mode: for example, in \figurename~\ref{fig:legGFL}, the first color in the right list represents a mode where grid currents, \ac{vsc} currents, and \ac{vsc} voltages interact.
\acp{pf} are displayed in a matrix, with columns representing modes and rows representing state variables: black denotes maximum participation, white denotes none \cite{linear1}. 

\subsection{Grid-Following Converter}
\begin{figure}[t!] 
    \centering
    \includegraphics[width=.95\linewidth]{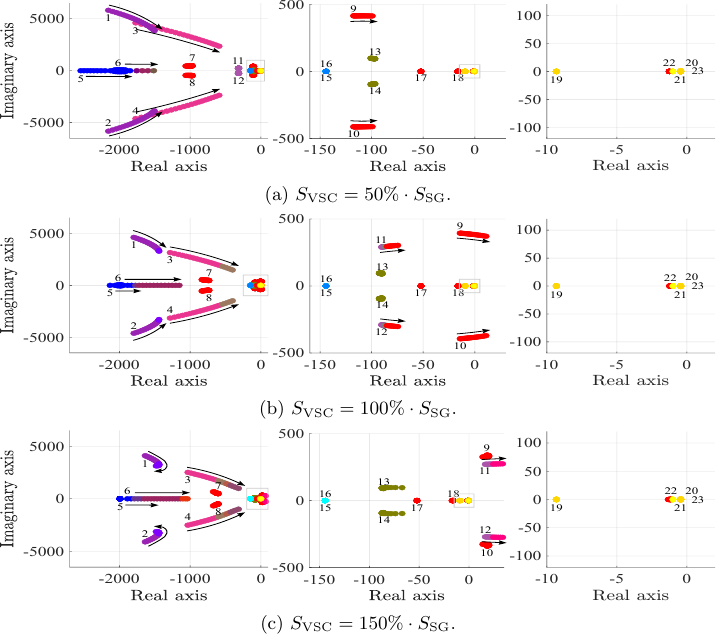}
    \caption{Eigenvalue trajectories of the system with \ac{gfl} \ac{vsc} when \ac{scr} is varied from $3.5$ to $1.5$, as indicated by the arrows.}
    \label{fig:eigGFL}
\end{figure}

\begin{figure}[t!]
    \centering
    \includegraphics[width=0.68\linewidth]{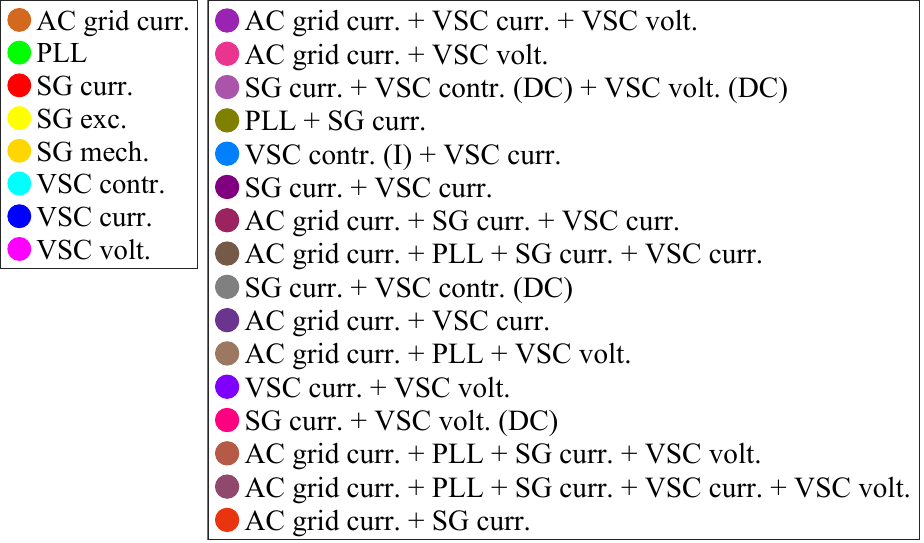}
    \caption{Pole color based on state variables with \acp{pf} $>30\%$ (\ac{gfl}).}
    \label{fig:legGFL}
\end{figure}

The eigenvalue trajectories of the system with the GFL converter are plotted in \figurename~\ref{fig:eigGFL}, considering the three \ac{vsc} power ratings and varying the \ac{scr} from $3.5$ to $1.5$; the corresponding pole color legend is shown in \figurename~\ref{fig:legGFL}.

When the \ac{vsc} operates at half \ac{sg} power, the system maintains stability for each \ac{scr} tested: the load is supplied mainly by the \ac{sg} and the voltage drop across the connection line is not high enough to cause \ac{gfl} instability in a weak-grid scenario. Conversely, if the \ac{vsc} and \ac{sg} powers are equal, the system is unstable for \ac{scr} $<2.2$ due to poles 9-10 with high participation of \ac{sg} currents in \figurename~\ref{fig:eigGFL}b. Also, other interactions may contribute to instability: the DC voltage control stops to interact with \ac{sg} (poles 11-12 change color as SCR is varied) as the grid becomes weaker, and the \ac{pll} interacts with \ac{sg} (poles 13-14). In this case, the load is mainly supplied by the \ac{vsc} and the voltage drop across the connection line increases, thereby increasing the \ac{vsc} injected current; in weak grid conditions, this results in significant \ac{pcc} voltage variations, negatively affecting DC-link dynamics, eventually leading to \ac{pll} loss of synchronism. The conditions become more critical when the \ac{vsc} power exceeds the \ac{sg} power, as the system is always unstable due to poles 9-10 and 11-12, as shown in \figurename~\ref{fig:eigGFL}c: most of the power absorbed by the load comes from the \ac{vsc} which thus cannot maintain its proper operation. 

\begin{figure}[t!]
    \centering
    \includegraphics[width=1\linewidth]{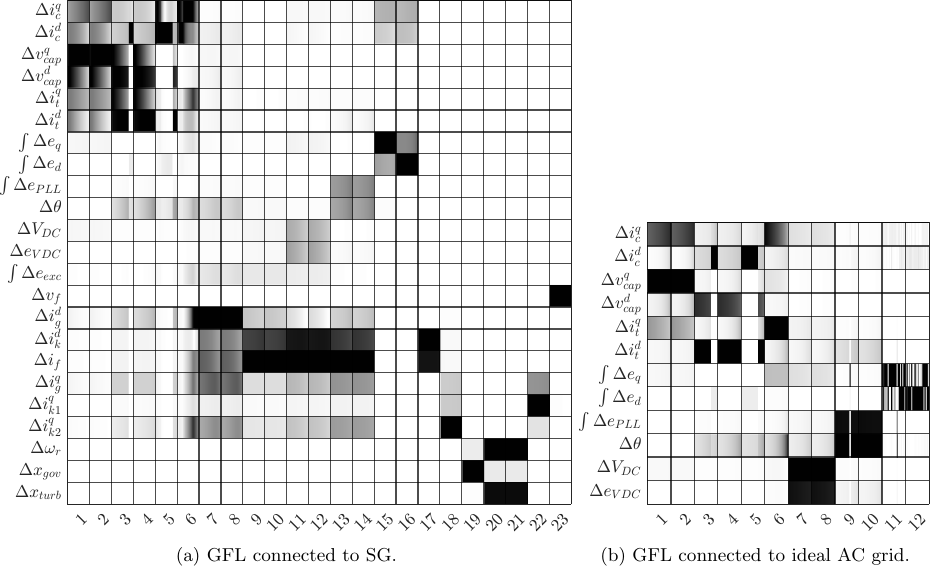}
    \caption{\acp{pf} of the system with \ac{gfl} \ac{vsc} for \ac{scr} varying from 3.5 to 1.5 and $S_\mathrm{VSC} = S_\mathrm{SG}$; each vertical band corresponds to the \ac{scr} values.}
    \label{fig:PFGFL}
\end{figure}

Furthermore, to understand the limitations of stability studies when considering an ideal AC grid, the same analyses are repeated by replacing the \ac{sg}-based power plant with an ideal AC grid. First, the \ac{scr} stability boundaries change significantly: considering, for example, $S_\mathrm{VSC} = S_\mathrm{grid}$, the system becomes unstable for \ac{scr} $< 1$. The \acp{pf} shown in \figurename~\ref{fig:PFGFL} reveal that, differently from the case with the \ac{sg}, the \ac{pll} and the DC voltage control do not interact with currents (modes associated with pole pairs 9-10 and 7-8 in \figurename~\ref{fig:PFGFL}b); instead, the angle difference between the two \acp{rf} still couples with grid currents, leading to instability for low \ac{scr} values: modes 4, 5, 6 in \figurename~\ref{fig:PFGFL}b show high participation of grid currents and $\Delta\theta$ as the \ac{scr} decreases. It is now evident that the ideal AC grid-based analysis exhibits a wider stability range compared to that with the real power grid and neglects the coupling effects between the \ac{gfl} and the grid dynamics: ignoring the frequency and voltage dynamics of an \ac{sg}-based power grid may lead to inaccurate evaluation of \ac{gfl}-related oscillatory modes and consequently of its stability margins.
\subsection{Grid-Forming Converter}
\begin{figure}[t!]
    \centering
    \includegraphics[width=.95\linewidth]{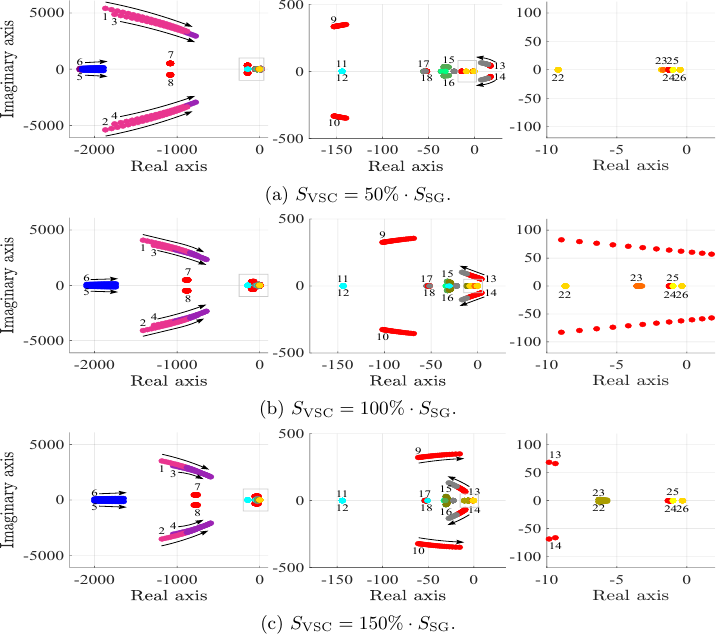}
    \caption{Eigenvalue trajectories of the system with \ac{gfm} \ac{vsc} when \ac{scr} is varied from $3.5$ to $1.5$, as indicated by the arrows.}
    \label{fig:eigGFM}
\end{figure}

\begin{figure}[t!]
    \centering
    \includegraphics[width=0.51\linewidth]{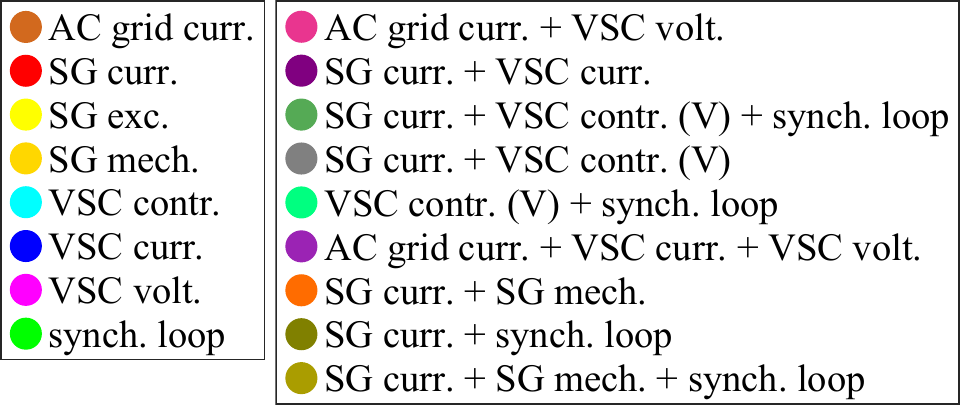}
    \caption{Pole color based on state variables with \acp{pf} $>30\%$ (\ac{gfm}).}
    \label{fig:legGFM}
\end{figure}

The same analysis is performed with the \ac{gfm} converter and the eigenvalue trajectories are plotted in \figurename~\ref{fig:eigGFM}; the corresponding pole color legend is shown in \figurename~\ref{fig:legGFM}.

If the \ac{vsc} power is half the \ac{sg} power, that is, the load is mostly supplied by the \ac{sg}, the system is always unstable due to poles 13-14 which represent an interaction between \ac{sg} currents and \ac{gfm} voltage control, as shown in \figurename~\ref{fig:eigGFM}a. In this case, since the \ac{vsc} injects less power than the \ac{sg}, 
the small phase difference between \ac{pcc} and \ac{sg} voltages may cause critical active power fluctuations, thus \ac{gfm} loss of synchronism and instability. 
When the \ac{vsc} matches the \ac{sg} power, the system, initially unstable, becomes stable in weak-grid scenarios (\ac{scr} $\le 2.4$). As previously noted, the instability is caused by the interaction between \ac{sg} currents and \ac{gfm} voltage control (poles 13-14) and by the coupling between \ac{sg} currents and synchronization loop (poles 15-16), as shown in \figurename~\ref{fig:eigGFM}b. In a strong power grid, the \ac{gfm} converter and the \ac{sg} are electrically close to each other, making the \ac{gfm} voltage regulation and the active power-based synchronization more challenging. 
Lastly, if the \ac{vsc} power exceeds the \ac{sg} power, the system is stable for each \ac{scr} tested: the load is mainly supplied by the \ac{gfm} converter, whose dynamics therefore prevails. In particular, the \ac{sg} frequency is coupled with \ac{sg} electrical and \ac{gfm} synchronization loop state variables and not only with governor and turbine variables (poles 22-23 in \figurename~\ref{fig:eigGFM}c change color with respect to the previous cases). 

\begin{figure}[t!]
    \centering
    \includegraphics[width=1\linewidth]{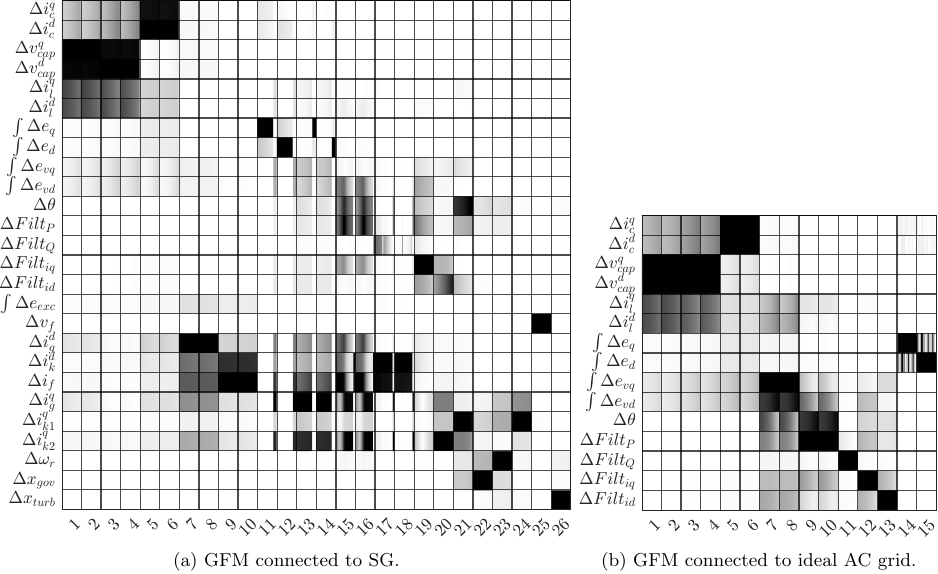}
    \caption{\acp{pf} of the system with \ac{gfm} \ac{vsc} for \ac{scr} varying from 3.5 to 1.5 and $S_\mathrm{VSC} = S_\mathrm{SG}$; each vertical band corresponds to the \ac{scr} values.}
    \label{fig:PFGFM}
\end{figure}

As before, substituting the \ac{sg} with an ideal grid, yields different stability margins: if $S_\mathrm{VSC} = S_\mathrm{grid}$, the system is stable for \ac{scr} $<2.1$, leading to a narrower stability range. Again, the instability is caused by the interaction between the \ac{gfm} voltage control and synchronization loop with the grid currents, as shown by the \acp{pf} in \figurename~\ref{fig:PFGFM}b (poles 7-8). However, when considering an ideal AC grid, the coupling between the grid frequency and currents cannot be captured. Indeed, poles 22-23 in \figurename~\ref{fig:PFGFM}a indicate that the \ac{sg} speed couples with both frequency regulation system and \ac{sg} currents: the \ac{sg} frequency depends both on the mechanical torque and the \ac{vsc} power, meaning that the \ac{vsc} may be able to actively compensate for any load variation. This analysis suggests that considering an ideal grid in \ac{ssa} may prevent the observation of all \ac{gfm}-related oscillatory modes and dynamics.

\section{Simulation results}
In order to verify the validity of the developed linear model and the aforementioned \ac{ssa}, the case study with $S\ped{VSC} = S\ped{SG}$ is implemented in MATLAB-Simulink and time-domain simulations are performed applying 1\% step increment in the active power absorbed by the load at $t = 1.5$ s. 

\begin{figure}[b!]
    \centering
    \includegraphics[width=0.758\linewidth]{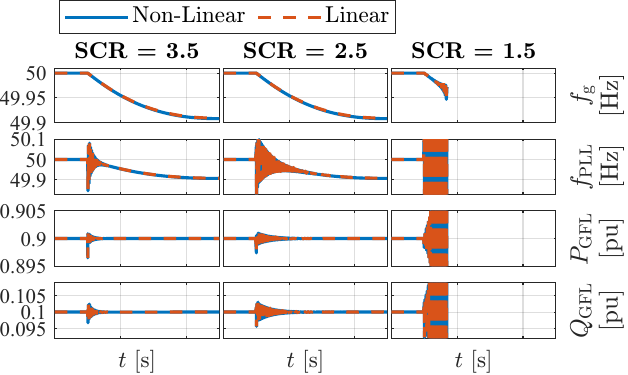}
    \caption{Simulation results with \ac{gfl} for different \ac{scr}.}
    \label{fig:valGFL}
\end{figure}

The \ac{sg} and \ac{pll} frequencies and the active and reactive powers injected by \ac{gfl} \ac{vsc} are shown in \figurename~\ref{fig:valGFL}. As the \ac{scr} decreases, the system damping decreases, leading to wider oscillations. When $\text{SCR} = 1.5$, the system is unstable and the simulation stops at $t = 1.82$ s; the increasing-amplitude oscillations are caused by the DC-link control and the \ac{pll}, leading to instability due to the negative damping of oscillatory mode associated with pole pair 9-10 ($f = 62$ Hz).

\begin{figure}[b!]
    \centering
    \includegraphics[width=0.758\linewidth]{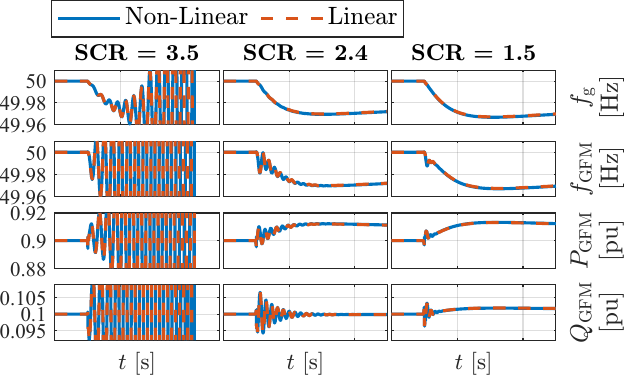}
    \caption{Simulation results with \ac{gfm} for different \ac{scr}.}
    \label{fig:valGFM}
\end{figure}

Similarly, the \ac{sg} and \ac{gfm} frequencies and the active and reactive powers injected by \ac{gfm} \ac{vsc} are shown in \figurename~\ref{fig:valGFM}. When $\text{SCR} = 3.5$, the low-frequency oscillations are caused by \ac{vsc} voltage control and then transferred to the system through the active power-based synchronization loop, causing oscillatory instability ($f = 10$ Hz) due to negative damping of the oscillatory mode associated with poles 13-14. Lastly, the system response becomes slower and better damped when the \ac{scr} decreases from $2.4$ to $1.5$. 
\section{Conclusion}
\label{sec5}
This work provides an analysis of the interactions between traditional \ac{sg}-based power plants and \acp{vsc}, relying on conventional \ac{gfl} and \ac{gfm} control schemes. Different case studies have been conducted by varying the converter rating power and the grid \ac{scr}, revealing that the stability of \acp{vsc}-dominated grids is strongly influenced by both and highly depends on the modeling of the power grid dynamics.

The \ac{gfl} converter operates correctly if its power is lower than the \ac{sg} power or in strong-grid scenarios. In weak-grid scenarios and when the \ac{gfl} power is considerably greater than the \ac{sg} power, the system is unstable due to DC-link dynamics and \ac{pll} loss of synchronism. 
In contrast, the \ac{gfm} converter experiences unstable operation when it is electrically close to the \ac{sg}; this occurs when its power is lower than \ac{sg} power or in strong-grid scenarios and is primarily due to the \ac{gfm} voltage control. When the \ac{gfm} power exceeds the \ac{sg} power, the converter dynamics prevails, the \ac{sg} follows accordingly, and the system is always stable. 

Consequently, the limitations of assuming an ideal power grid instead of a real \ac{sg}-based one have been investigated: compared to the real grid, the ideal grid-based analysis exhibits wider \ac{gfl} and narrower \ac{gfm} stability ranges. Moreover, some important coupling effects are neglected: in the case of \ac{gfl} control, the DC voltage dynamics and \ac{pll} no longer interact with grid currents, while, in the case of \ac{gfm} control, the grid frequency no longer interacts with grid currents.

Lastly, the findings from the small-signal analysis have been validated with time-domain simulations. 

\bibliographystyle{IEEEtran} 

\end{document}